\PassOptionsToPackage{unicode}{hyperref}
\PassOptionsToPackage{hyphens}{url}
\documentclass[
]{article}
\usepackage{xcolor}
\usepackage[margin=1in]{geometry}
\usepackage{amsmath,amssymb}
\usepackage{iftex}
\ifPDFTeX
  \usepackage[T1]{fontenc}
  \usepackage[utf8]{inputenc}
  \usepackage{textcomp} 
\else 
  \usepackage{unicode-math} 
  \defaultfontfeatures{Scale=MatchLowercase}
  \defaultfontfeatures[\rmfamily]{Ligatures=TeX,Scale=1}
\fi
\usepackage{lmodern}
\ifPDFTeX\else
\fi
\IfFileExists{upquote.sty}{\usepackage{upquote}}{}
\IfFileExists{microtype.sty}{
  \usepackage[]{microtype}
  \UseMicrotypeSet[protrusion]{basicmath} 
}{}
\makeatletter
\@ifundefined{KOMAClassName}{
  \IfFileExists{parskip.sty}{%
    \usepackage{parskip}
  }{
    \setlength{\parindent}{0pt}
    \setlength{\parskip}{6pt plus 2pt minus 1pt}}
}{
  \KOMAoptions{parskip=half}}
\makeatother
\usepackage{graphicx}
\makeatletter
\newsavebox\pandoc@box
\newcommand*\pandocbounded[1]{
  \sbox\pandoc@box{#1}%
  \Gscale@div\@tempa{\textheight}{\dimexpr\ht\pandoc@box+\dp\pandoc@box\relax}%
  \Gscale@div\@tempb{\linewidth}{\wd\pandoc@box}%
  \ifdim\@tempb\p@<\@tempa\p@\let\@tempa\@tempb\fi
  \ifdim\@tempa\p@<\p@\scalebox{\@tempa}{\usebox\pandoc@box}%
  \else\usebox{\pandoc@box}%
  \fi%
}
\def\fps@figure{htbp}
\makeatother
\NewDocumentCommand\citeproctext{}{}

\makeatletter
 \let\@cite@ofmt\@firstofone
 \def\@biblabel#1{}
 \def\@cite#1#2{{#1\if@tempswa , #2\fi}}
\makeatother
\newlength{\cslhangindent}
\newlength{\csllabelwidth}
\newenvironment{CSLReferences}[2] 
 {\begin{list}{}{%
  \setlength{\itemindent}{0pt}
  \setlength{\leftmargin}{0pt}
  \setlength{\parsep}{0pt}
  \ifodd #1
   \setlength{\leftmargin}{\cslhangindent}
   \setlength{\itemindent}{-1\cslhangindent}
  \fi
  \setlength{\itemsep}{#2\baselineskip}}}
 {\end{list}}
\usepackage{calc}

\providecommand{\tightlist}{%
  \setlength{\itemsep}{0pt}\setlength{\parskip}{0pt}}
\usepackage{bookmark}
\IfFileExists{xurl.sty}{\usepackage{xurl}}{} 
\hypersetup{
  pdftitle={SimOmics: A Simulation Toolkit for Multivariate and Multi-Omics Data},
  pdfauthor={Kaitao Lai1},
  hidelinks,
  pdfcreator={LaTeX via pandoc}}

\title{SimOmics: A Simulation Toolkit for Multivariate and Multi-Omics
Data}
\author{Kaitao Lai\textsuperscript{1}}
\date{8 July 2025}

\begin{document}
\maketitle

\textsuperscript{1} University of Sydney

\section{Summary}\label{summary}

SimOmics is an R package designed to generate realistic, multivariate,
and multi-omics synthetic datasets. It is intended for use in
benchmarking, method development, and reproducibility in bioinformatics,
particularly in the context of omics integration tasks such as those
encountered in transcriptomics, proteomics, and metabolomics. SimOmics
supports latent factor simulation, sparsity structures, block-wise
covariance modeling, and biologically inspired noise models and feature
dimensions.

\section{Statement of need}\label{statement-of-need}

Simulated datasets play a crucial role in statistical method
development, especially in fields such as systems biology and
multi-omics integration, where complex relationships often exist between
multiple data modalities. However, existing simulation tools are either
too simple (e.g., based on basic multivariate distributions) or lack
support for biologically relevant structures such as sparsity, block
integration, or shared latent variables.

SimOmics fills this gap by providing an accessible yet powerful
framework to generate synthetic multi-omics datasets that resemble real
biological complexity. This is critical for the testing and benchmarking
of methods like mixOmics (Rohart et al. 2017), MOFA2 (Argelaguet et al.
2020), and iCluster (Wang et al. 2014).

\section{Features}\label{features}

\begin{itemize}
\tightlist
\item
  Simulate multiple omics blocks with flexible dimensions
\item
  Inject shared latent factors or independent noise
\item
  Control inter-block correlation via a block covariance structure
\item
  Add Gaussian noise and customize signal-to-noise ratio
\item
  Plot PCA, correlation, and integration results
\item
  Export data and integrate with other packages like \texttt{mixOmics}
\end{itemize}

\section{Why Simulated Data Matters}\label{why-simulated-data-matters}

Simulated datasets---when designed to reflect realistic biological
complexity---are essential tools in bioinformatics research (Zhang et
al. 2020; Huang et al. 2017). SimOmics enables users to generate such
datasets for a range of multivariate and multi-omics scenarios.

\subsection{Advantages of synthetic
datasets:}\label{advantages-of-synthetic-datasets}

\begin{itemize}
\tightlist
\item
  \textbf{Method development}: Early-stage algorithm development
  benefits from controllable data without needing access to real
  datasets (Witten, Tibshirani, and Hastie 2009).
\item
  \textbf{Known ground truth}: Simulated data allow researchers to
  benchmark models based on sensitivity, specificity, and overall
  performance with known signal.
\item
  \textbf{Stress-testing}: Users can design edge-case scenarios (e.g.,
  high noise, low correlation, small sample size) not typically found in
  public datasets (Teschendorff and Relton 2018).
\item
  \textbf{Reproducibility}: Synthetic data ensures that benchmarking can
  be reproduced across institutions and software implementations.
\end{itemize}

\section{Use cases and target
audience}\label{use-cases-and-target-audience}

SimOmics is intended for:

\begin{itemize}
\tightlist
\item
  Researchers developing new multi-omics integration methods
\item
  Developers of machine learning models for high-dimensional data
\item
  Authors writing benchmarking papers that require known ground truth
\item
  Bioinformatics instructors demonstrating omics integration techniques
\end{itemize}

Users include PhD students, statisticians, and software developers in
computational biology. Although it is not aimed at experimental
biologists, it is a valuable tool for method validation and reproducible
research.

\section{Related work}\label{related-work}

Several existing R packages such as \texttt{clusterGeneration},
\texttt{bnlearn}, and \texttt{simuPOP} support general-purpose
simulation of data. However, none offer block-wise omics simulation with
latent factors and realistic biological noise models. SimOmics builds on
the need observed in published tools such as mixOmics (Rohart et al.
2017), MOFA2 (Argelaguet et al. 2020), and iCluster (Wang et al. 2014)
to benchmark model performance in structured settings.

\section{Example Use Case}\label{example-use-case}

The following figure shows the result of applying
\texttt{mixOmics::block.plsda()} to two synthetic omics blocks
(transcriptome and proteome) generated by SimOmics. Class labels A and B
are partially separable, reflecting shared latent structure while
preserving variability.

\begin{figure}
\centering
\pandocbounded{\includegraphics[keepaspectratio,alt={PLS-DA of simulated multi-omics dataset}]{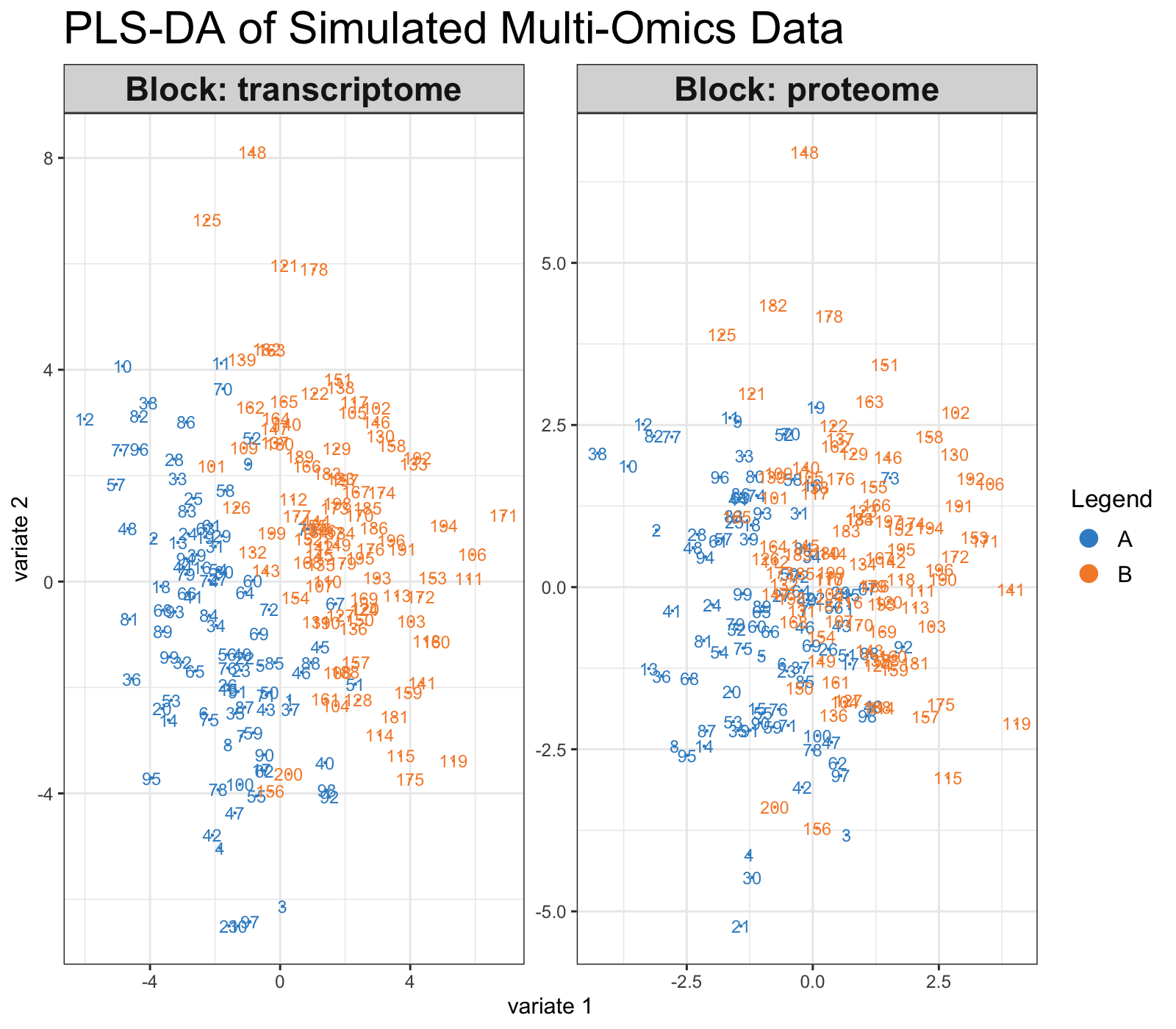}}
\caption{PLS-DA of simulated multi-omics dataset}
\end{figure}

\textbf{Figure 1.} PLS-DA of the simulated multi-omics dataset. Samples
(colored by class A and B) are projected into a shared latent space. The
partial overlap illustrates integrated but not fully
class-discriminative structure --- a common challenge in real-world
omics data.

\section{Software Repository}\label{software-repository}

The source code for SimOmics is freely available on GitHub at:\\
\url{https://github.com/biosciences/SimOmics}

\section{Acknowledgements}\label{acknowledgements}

This project was initiated to support method development and
reproducible benchmarking in multi-omics research. We acknowledge the
contributions of the broader open-source and mixOmics communities.

\section*{References}\label{references}
\addcontentsline{toc}{section}{References}

\protect\phantomsection\label{refs}


\begin{CSLReferences}{1}{0}
\bibitem[\citeproctext]{ref-argelaguet2020mofa2}
Argelaguet, Ricard, Denis Arnol, Dmitry Bredikhin, Yann Deloro, Lars
Velten, John C Marioni, Florian Buettner, Wolfgang Huber, and Oliver
Stegle. 2020. {``MOFA+: A Statistical Framework for Comprehensive
Integration of Multi-Modal Single-Cell Data.''} \emph{Genome Biology}
21: 111. \url{https://doi.org/10.1186/s13059-020-02015-1}.

\bibitem[\citeproctext]{ref-huang2017systematic}
Huang, Jing, Guoliang Zhou, Shujie Meng, Yan Chen, Tingting Dong, Zhi
Cheng, and Yu Wang. 2017. {``Systematic Evaluation of Molecular Networks
for Discovery of Disease Genes.''} \emph{Cell Systems} 5 (5): 460--70.
\url{https://doi.org/10.1016/j.cels.2017.09.006}.

\bibitem[\citeproctext]{ref-rohart2017mixomics}
Rohart, Florian, Benoit Gautier, Amrit Singh, and Kim-Anh Lê Cao. 2017.
{``mixOmics: An r Package for 'Omics Feature Selection and Multiple Data
Integration.''} \emph{PLoS Computational Biology} 13 (11): e1005752.
\url{https://doi.org/10.1371/journal.pcbi.1005752}.

\bibitem[\citeproctext]{ref-teschendorff2018statistical}
Teschendorff, Andrew E, and Caroline L Relton. 2018. {``Statistical and
Integrative System-Level Analysis of DNA Methylation Data.''}
\emph{Nature Reviews Genetics} 19 (3): 129--47.
\url{https://doi.org/10.1038/nrg.2017.86}.

\bibitem[\citeproctext]{ref-wang2014iclatplus}
Wang, Bo, Asif M Mezlini, Fahad Demir, Marc Fiume, Zhaleh Tu, Michael
Brudno, Benjamin Haibe-Kains, and Anna Goldenberg. 2014.
{``iClusterPlus: Integrative Clustering of Multiple Genomic Data Types
with Feature Selection.''} \emph{Statistical Applications in Genetics
and Molecular Biology} 13 (6): 511--26.
\url{https://doi.org/10.1515/sagmb-2013-0059}.

\bibitem[\citeproctext]{ref-witten2009penalized}
Witten, Daniela M, Robert Tibshirani, and Trevor Hastie. 2009. {``A
Penalized Matrix Decomposition, with Applications to Sparse Principal
Components and Canonical Correlation Analysis.''} \emph{Biostatistics}
10 (3): 515--34. \url{https://doi.org/10.1093/biostatistics/kxp008}.

\bibitem[\citeproctext]{ref-zhang2020synthetic}
Zhang, Honghan, Yi Wang, Qi Wang, Wenjie Zheng, Zhaonan Huang, and Zhen
Huang. 2020. {``Synthetic Data Generation and Its Application in Machine
Learning for Healthcare.''} \emph{Npj Digital Medicine} 3: 106.
\url{https://doi.org/10.1038/s41746-020-00343-7}.

\end{CSLReferences}
\end{document}